\documentclass[12pt,preprint]{aastex}
\usepackage{natbib,makeidx}
\shorttitle{The 2005 outburst of XTE J1118+480}
\shortauthors{Zurita et al.}

\begin{document}


\title{The 2005 outburst of the halo black hole X-ray transient XTE J1118+480}


\author{C. Zurita}
\affil{Instituto de Astrof\'{\i}sica de Canarias, 38200 La Laguna, Tenerife, Spain}
\email{czurita@iac.es}

\author{M.A.P. Torres}
\affil{Harvard-Smithsonian Center for Astrophysics, 60 Garden St, Cambridge, MA 02138, USA}
%\email{mtorres@head.cfa.harvard.edu}

\author{D.Steeghs}
\affil{Harvard-Smithsonian Center for Astrophysics, 60 Garden St, Cambridge, MA 02138, USA}
%\email{}

\author{P. Rodr\'{\i}guez-Gil}
\affil{Instituto de Astrof\'{\i}sica de Canarias, 38200 La Laguna, Tenerife, Spain}
%\email{prguez@iac.es}

\author{T. Mu\~noz-Darias}
\affil{Instituto de Astrof\'{\i}sica de Canarias, 38200 La Laguna, Tenerife, Spain}
%\email{tmd@iac.es}

\author{J. Casares}
\affil{Instituto de Astrof\'{\i}sica de Canarias, 38200 La Laguna, Tenerife, Spain}
%\email{jcv@iac.es}

\author{T. Shahbaz}
\affil{Instituto de Astrof\'{\i}sica de Canarias, 38200 La Laguna, Tenerife, Spain}
%\email{}

\author{I.G. Mart\'{\i}nez-Pais}
\affil{Instituto de Astrof\'{\i}sica de Canarias, 38200 La Laguna, Tenerife,
  Spain}
\affil{Departamento de Astrof\'{\i}sica, Universidad de La Laguna, E-38206 La
  Laguna, Tenerife, Spain}

\author{P. Zhao}
\affil{Harvard-Smithsonian Center for Astrophysics, 60 Garden St, Cambridge, MA 02138, USA}
%\email{}

\author{M.R. Garcia}
\affil{Harvard-Smithsonian Center for Astrophysics, 60 Garden St, Cambridge, MA 02138, USA}
%\email{}

\author{A. Piccioni}
\affil{Dipartimento di Astronomia, Universit\'a di Bologna, Via Ranzani 1, Bologna, Italy}
%\email{piccioni@astbo4.bo.astro.it}

\author{C. Bartolini}
\affil{Dipartimento di Astronomia, Universit\'a di Bologna, Via Ranzani 1, Bologna, Italy}
%\email{piccioni@astbo4.bo.astro.it}

\author{A. Guarnieri}
\affil{Dipartimento di Astronomia, Universit\'a di Bologna, Via Ranzani 1, Bologna, Italy}

\author{J.S. Bloom}
\affil{Astronomy Department, University of California at Berkeley, Berkeley, CA 94720, USA}

\author{C.H. Blake}
\affil{Harvard-Smithsonian Center for Astrophysics, 60 Garden St, Cambridge, MA 02138, USA}

\author{E.E.Falco}
\affil{Harvard-Smithsonian Center for Astrophysics, 60 Garden St, Cambridge, MA 02138, USA}
\author{A.Szentgyorgyi}
\affil{Harvard-Smithsonian Center for Astrophysics, 60 Garden St, Cambridge, MA 02138, USA}
\author{M.Skrutskie}
\affil{University of Virginia, Department of Astronomy, PO Box 3818, 
Charlottesville, VA 22903, USA}
%\email{piccioni@astbo4.bo.astro.it}

%\email{}


\begin{abstract}
We present optical  and infrared monitoring of the 2005  outburst of the halo
black  hole X-ray  transient XTE  J1118+480.   We measured  a total  outburst
amplitude of  $\sim5.7\pm0.1$\,mag in  the $R$ band  and $\sim5$\,mag  in the
infrared  $J$,  $H$  and   $K_\mathrm{s}$  bands.   The  hardness  ratio  HR2
(5-12\,keV/3-5 keV)  from the RXTE/ASM data  is 1.53$\pm$0.02 at  the peak of
the outburst indicating  a hard spectrum.  Both the shape  of the light curve
and   the  ratio   $L_\mathrm{X}$(1-10\,keV)$/L_\mathrm{opt}$   resemble  the
mini-outbursts  observed in  GRO J0422+32  and XTE  J1859+226.   During early
decline, we find  a 0.02-mag amplitude variation consistent  with a superhump
modulation, like  the one observed  during the 2000 outburst.  Similarly, XTE
J1118+480  displayed a  double-humped ellipsoidal  modulation distorted  by a
superhump wave when settled into a near-quiescence level, suggesting that the
disk expanded  to the 3:1 resonance  radius after outburst  where it remained
until early  quiescence.  The  system reached quiescence  at $R=19.02\pm0.03$
about  three  months after  the  onset of  the  outburst.   The optical  rise
preceded the X-ray rise by at most 4 days.  The spectral energy distributions
(SEDs) at the different epochs  during outburst are all quasi-power laws with
$F_{\nu}\propto\nu^{\alpha}$ increasing toward the  blue.  At the peak of the
outburst  we derived  $\alpha=0.49\pm0.04$  for the  optical  data alone  and
$\alpha=0.1\pm0.1$ when fitting solely the infrared.  This difference between
the optical and the infrared SEDs  suggests that the infrared is dominated by
a different component (a jet?)  whereas the optical is presumably showing the
disk evolution.
\end{abstract}


%% subject header.  Objects should be in the appropriate "individual"
%% headers (e.g. quasars: individual, stars: individual, etc.) with the
%% additional provision that the total number of headers, including each
%% individual object, not exceed six.  The \objectname{} macro, and its
%% alias \object{}, is used to mark each object.  The macro takes the object
%% name as its primary argument.  This name will appear in the paper
%% and serve as the link's anchor in the electronic edition if the name
%% is recognized by the data centers.  The macro also takes an optional
%% argument in parentheses in cases where the data center identification
%% differs from what is to be printed in the paper.

\keywords{accretion, accretion disks ---
binaries: close --- stars:individual (XTE J1118+480, KV UMa)--- X-rays: stars}

\section{Introduction}

X-ray transients (XRTs) are a class  of low-mass X-ray binaries in which long
periods  of  quiescence  (typically  decades)  are  interrupted  by  dramatic
outbursts, when the X-ray luminosity suddenly  increases by up to a factor of
10$^6$ \citep[e.g.,][]{charles04}.  During the outburst, XRTs usually reach a
state in which  the X-ray emission is dominated by  thermal emission from the
hot inner accretion disk (i.e.  the  {\em high/soft} -- HS -- or {\em thermal
dominated} state).   There are,  however, a few  of these  systems \citep[see
e.g.,][]{brocksopp04}  that  are  instead  dominated by  a  hard  non-thermal
power-law component likely produced by thermal Comptonization of seed photons
in the vicinity  of the accreting black hole (i.e.  the  {\em low/hard} -- LH
-- state).  At even lower accretion rates, XRTs reach quiescence which may be
just an  extreme example of the LH  state. To explain the  LH (and quiescent)
state a standard  disk truncated at some large inner  radius is assumed.  The
interior  volume  is  filled  with  a hot,  optically  thin,  quasi-spherical
accretion  flow where  most of  the energy  released via  viscous dissipation
remains in  this flow rather than  being radiated away  (as in a disk)  to be
finally  advected  by the  compact  object \citep[e.g.,][]{narayan96}.   This
model, called Advection  Dominated Accretion Flow (ADAF), is  the most widely
discussed picture  although other alternatives  have also been  invoked.  For
instance, the Accretion  Disk Corona model assumes a  cool thin disk embedded
in a hot corona powered by magnetic flares \citep[e.g.,][]{merloni01}.  Also,
it  has been  proposed that  emission  from jets  (which are  believed to  be
associated with  the LH state) can  account for the observed  spectra in XRTs
\citep[e.g.,][]{merloni02}.\\

The XRT  XTE J1118+480 was discovered by  the All Sky Monitor  ({\sc asm}) on
board  of the  Rossi X-ray  Timing  Explorer ({\sc  rxte}) on  2000 March  29
\citep{remillard00}  as a  weak, slowly  rising X-ray  source.  Retrospective
analysis of  the {\sc asm} database  revealed a previous  outburst episode in
January 2000.  The precursor was shorter than the main outburst although both
reached similar brightness  (e.g. Uemura et al. 2000).   XTE J1118+480 is one
of  the few XRTs  that remained  in a  LH state  throughout the  outburst and
failed to  reach the  HS state.   This object is  also important  for several
reasons: it   is  a   secure  case   of  black   hole  \citep[$f(M)=6.1\pm0.3
M_\mathrm{\odot}$;][]{mcclintock01a,wagner01,torres04},    the   black   hole
transient        with        the        shortest        orbital        period
\citep[$P_\mathrm{orb}\sim4.1$\,hr;][]{patterson00,uemura00}   and  also  the
first  black hole  binary  in the  Galactic halo  \citep{wagner01,mirabel01}.
Additionally, the ratio $L_{\rm x}/L_{\rm opt} \simeq 5$ was extremely low so
it has  been suggested  that XTE  J1118+480 may be  an Accretion  Disk Corona
source seen  at high inclination  \citep{garcia00}.  However, no  eclipse has
ever   been  recorded  and   consequently,  the   source  should   have  been
intrinsically faint in X-rays  \citep{hynes00} during the outburst.  Finally,
associated radio emission from this source has been reported \citep{pooley00}
with jet interpretation \citep{fender01}.\\

The  low interstellar  absorption towards  XTE J1118+480  allowed  a detailed
multi-wavelength study during the 2000 outburst and quiescence being the only
XRT   for  which   extreme   ultraviolet  observations   could  be   achieved
\citep{hynes00,chaty03,mcclintock03}.  In  fact, the best  observations of an
XRT in the LH state have been  made for this system (together with GX 339-4).
These observations have  been of great significance in the  effort to build a
physical model  of the accretion flow.   XTE J1118+480 has  been described in
terms  of an  ADAF \citep{mcclintock01b,  esin01}, an  Accretion  Disk Corona
model         \citep{merloni01b}        and        a         jet        model
\citep{markoff01,malzac04,yuan05b}.\\

Here we report the  follow-up of a new outburst of the  halo black hole X-ray
transient XTE J1118+480 (hereafter J1118) and investigate the implications in
modeling the structure of accretion  flows onto black holes and the mechanisms
involved.\\

\section{Observations and reductions}

Optical photometry  was obtained with the  0.8\,m IAC80 and  the 1\,m Optical
Ground Station  (OGS) telescopes at  the Observatorio del Teide  on Tenerife,
the  1.52\,m telescope  at  Loiano,  the 1.2\,m  telescope  at Fred  Lawrence
Whipple Observatory (FLWO), and the  4.2\,m WHT at the Observatorio del Roque
de  los Muchachos  on La  Palma. The  target was  mainly imaged  in  $R$ band
although we also obtained some $B,V,I$ colors.  Integration times ranged from
10~s to 15~min depending on telescope size, atmospheric conditions and target
brightness.   All images  were corrected  for bias  and flat--fielded  in the
standard   way  using   IRAF\footnotemark\footnotetext{Image   Reduction  and
Analysis   Facility,   distributed   by   the  National   Optical   Astronomy
Observatories}.  We  performed aperture photometry on our  object and several
nearby  comparison  stars.  We  had  previously  performed a  color-dependent
calibration of  a set  of 14 stars  in the  7'$\times$7' field of  view using
several standard stars from 6 Landolt plates \citep{landolt92}.\\

Infrared      photometry      was      obtained     with      the      1.3\,m
PAIRITEL\footnotemark\footnotetext{\url{http://   www.pairitel.org}}  robotic
telescope at  FLWO.  The  camera is the  2MASS South instrument  which images
simultaneously in  $J$, $H$  and $K_\mathrm{s}$ covering  a field of  view of
8.5'$\times$8.5'.  A large number of dithered 7.8\,s exposures are first bias
and  flat-field  corrected  and  then  combined  to  form  mosaics  for  each
individual visit which is typically 600\,s long (see also Blake et al. 2005).
For each  visit, instrumental $JHK_\mathrm{s}$ magnitudes  were extracted and
consequently calibrated relative  to the same 5 nearby  2MASS sources for all
exposures. Photometric  error estimates on the  IR magnitudes are  based on a
combination  of Poisson  statistics and  the error  contribution of  the five
reference stars used for each observation.\\

The X-ray data were obtained from  the RXTE ASM public archive which contains
 several daily 2--12  keV scans of the source since  its first recorded X-ray
 outburst in 2000. Intensity measurements in three energy bands: 1.5--3, 3--5
 and 5--12\,keV were  also recorded. Before 2005 January  10, J1118 was never
 detected   above    the   10    mCrab   one-day   bin    sensitivity   level
 \citep{levine96}. After January  16, the available data either  do not yield
 good sensitivity measurements or they indicate  that the flux is 10 mCrab or
 less.\\

\section{Long term light curve}

\begin{figure}
\includegraphics[angle=0,scale=0.8]{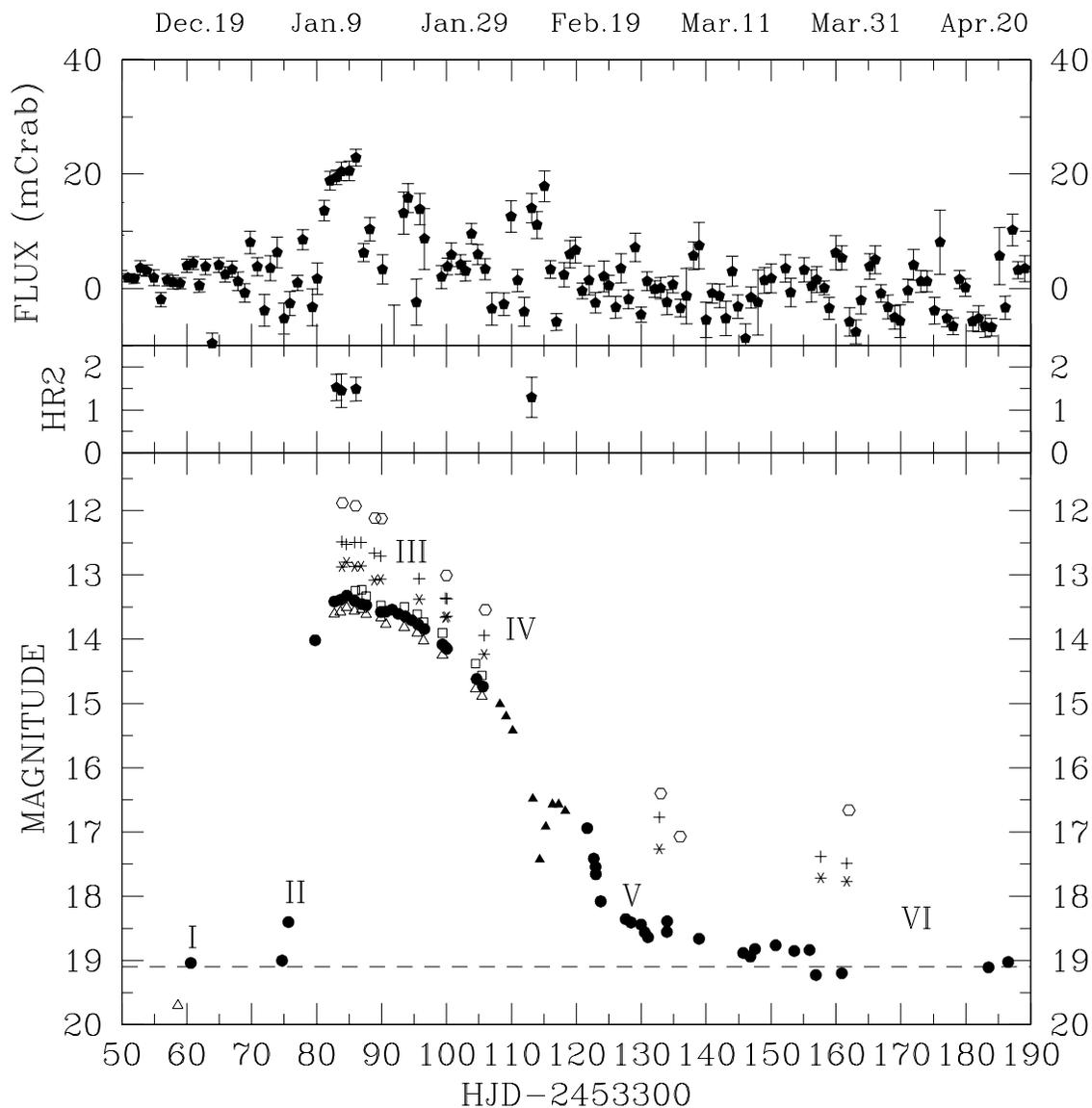}
\caption{ \label{outburstA} Long-term temporal  evolution of J1118 plotted as
  ({\em from top to bottom}): X-ray flux (the X-ray data were provided by the
  ASM/RXTE   teams  at   MIT  and   at  the   RXTE  SOF   at   NASA's  GSFC);
  (5-12\,keV)/(3-5\,keV) hardness ratio (HR2) derived from the ASM/RXTE data;
  Optical  magnitudes averaged  per  night where  optical  and infrared  data
  points  are: $B$ (diagonal  crosses), $V$  (triangles), $R$  (circles), $I$
  (squares),  $J$ (asterisks), $H$  (crosses) and  $K_\mathrm{s}$ (hexagons).
  Filled   triangles    mark   the   $V$   band    magnitudes   reported   by
  \citet{chou05}. The dashed  line marks the mean quiescent  magnitude in the
  $R$ band.}
\end{figure}

We  discovered an  optical rebrightening  of J1118  $\sim$5\,years  after its
first reported outburst \citep{zurita05}. We  then embarked on a new campaign
of  systematic monitoring  of the  outburst light  curve in  the  optical and
infrared bands.   The source reached  $R=13.37\pm0.01$ at the  outburst peak,
5.7$\pm$0.01 mag above the mean quiescent level.  In the infrared we measured
total outburst amplitudes of  $4.9\pm0.1$, $5.0\pm0.1$ and $4.8\pm0.1$ mag in
$J$, $H$  and $K_\mathrm{s}$ respectively.  Although  the optical brightening
is similar to  that observed during the 2000 outburst,  the current event did
not power  the source above $\sim  25$ mCrab in X-rays.   Similarly, the 2000
outburst, although  also faint in X-rays,  reached a peak of  $\sim 45$ mCrab
\citep{wood01}.    In    Fig.~\ref{outburstA}   we   present    our   overall
optical/infrared  light  curve  of   the  2005  outburst.   The  light  curve
morphology  of  both   the  X-ray  and  the  optical   emission  is  not  the
``canonical''  FRED \citep[fast rise  and exponential  decay;][]{chen97}.  In
comparison, the  precursor X-ray  light curve for  the 2000 outburst  shows a
FRED and the  March outburst (X-ray and optical)  showed a plateau morphology
\citep[see  the 2000  outburst  light curves  in  e.g.][]{wren01}.  We  note,
however, that  neither the sources that remained  in a LH state,  nor a large
number of those which reach the  HS state are FRED-shaped.  We calculated the
hardness ratio HR2 (5-12\,keV/3-5 keV)  from the RXTE/ASM data. Although only
a  few  measurements  were  possible  due  to  statistical  limitations  (see
Fig.~\ref{outburstA}), we found HR2=1.53$\pm$0.02 at the peak of the outburst
(from HJD-2453300=  83 to 86).  This value  of HR2 is consistent  with a hard
spectrum supporting that J1118 likely remained in the LH state throughout the
outburst \citep[e.g.][]{mcclintock05}.\\

\begin{figure}
\includegraphics[angle=0,scale=0.8]{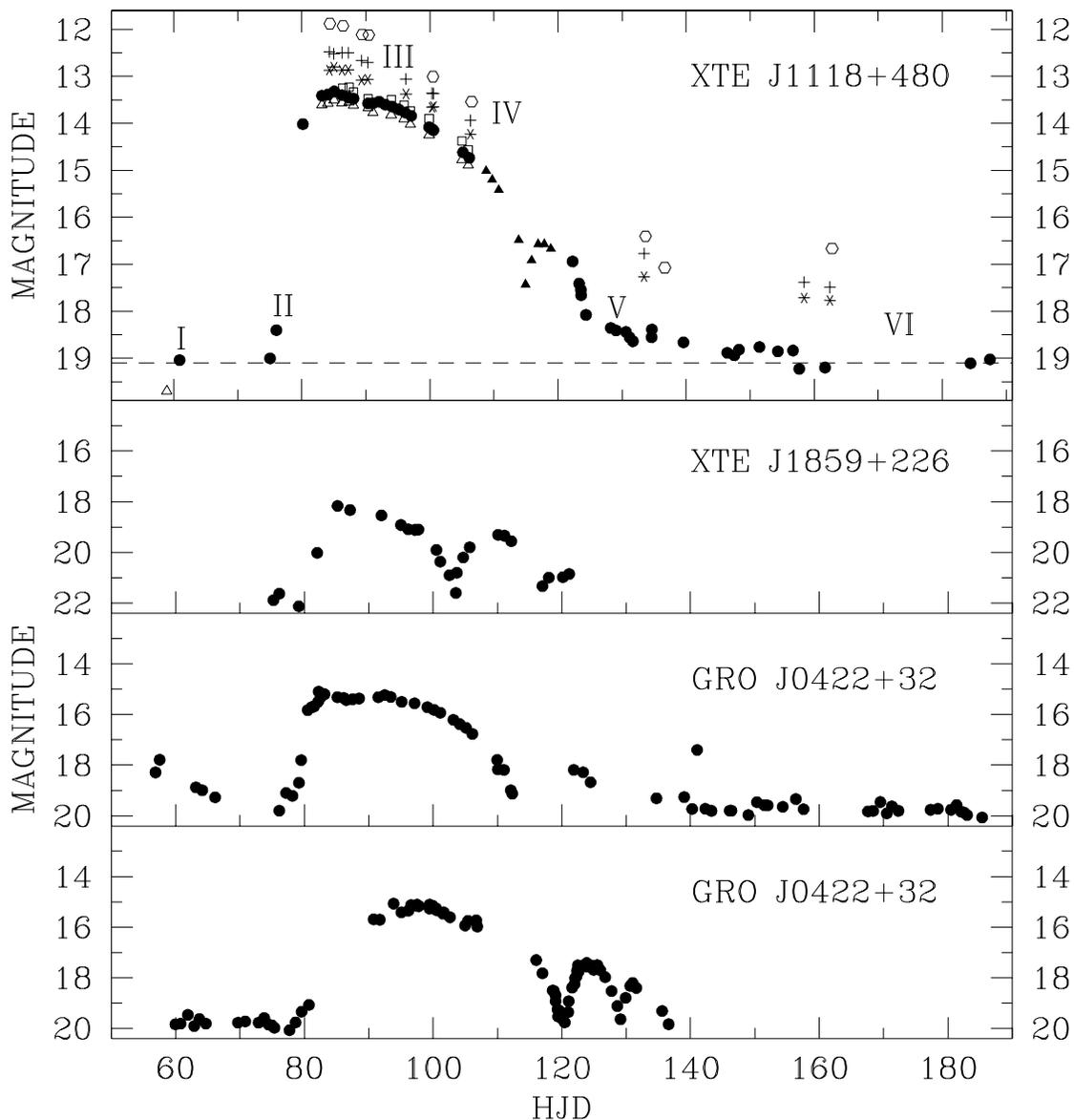}
\caption{ \label{outburstB}  {\em Top panel} Long-term  optical lightcurve of
  J1118.  Symbols for the optical and  infrared data points are the same than
  in Fig.~1.   {\em Bottom panels}:  The {\em mini-outburst}  $R$--band light
  curves  of   XTE  J1859+226  and   GRO  J0422+32  plotted   for  comparison
  \citep[complete   light  curves   of  these   systems  can   be   found  in
  ][]{zurita02b,callanan95}.  J1118 dates are HJD-2453300 while XTE J1859+226
  and GRO  J0422+32 curves  have been shifted  in time  to get the  same {\em
  mini-outburst}  starting  times.   The   J0422+32  data  were  provided  by
  E. Kuulkers.}
\end{figure}

The shape of  the 2005 outburst light curve is remarkably  similar to that of
the {\em mini-outbursts} observed in GRO J0422+32 \citep{chevalier95} and XTE
J1859+226  \citep{zurita02b}.  About 30  days after  the 2005  outburst peak,
there is evidence  for a short-lived re-flare with  amplitude of about 1\,mag
above  the pre-event  level and  lasting for  about 10  days  \citep[see also
filled  triangles in  Fig.~\ref{outburstA}]{chou05}.  Furthermore,  the light
curve displays  a small bump just  before the system  definitely settled down
into  quiescence.  This  type  of behavior  was  also found  during the  {\em
mini-outbursts}   of    both   GRO   J0422+32   and    XTE   J1859+226   (see
Fig.~\ref{outburstB}).   These  light   curves  have  comparable  length  and
brightness and also  the X-ray to optical ratio measured  in J1118 is similar
to what has been seen in {\em  mini-outbursts}, i.e. it is much lower than in
normal  outbursts   \citep[e.g.,][]{zurita02b}.   These  facts   support  the
suggestion of  \citet{hynes00} that  the outbursts in  J1118 are  indeed {\em
mini-outbursts} rather than full XRT events.

\section{Optical modulation during early decline and near quiescence}

\begin{figure}
\includegraphics[angle=0,scale=0.8]{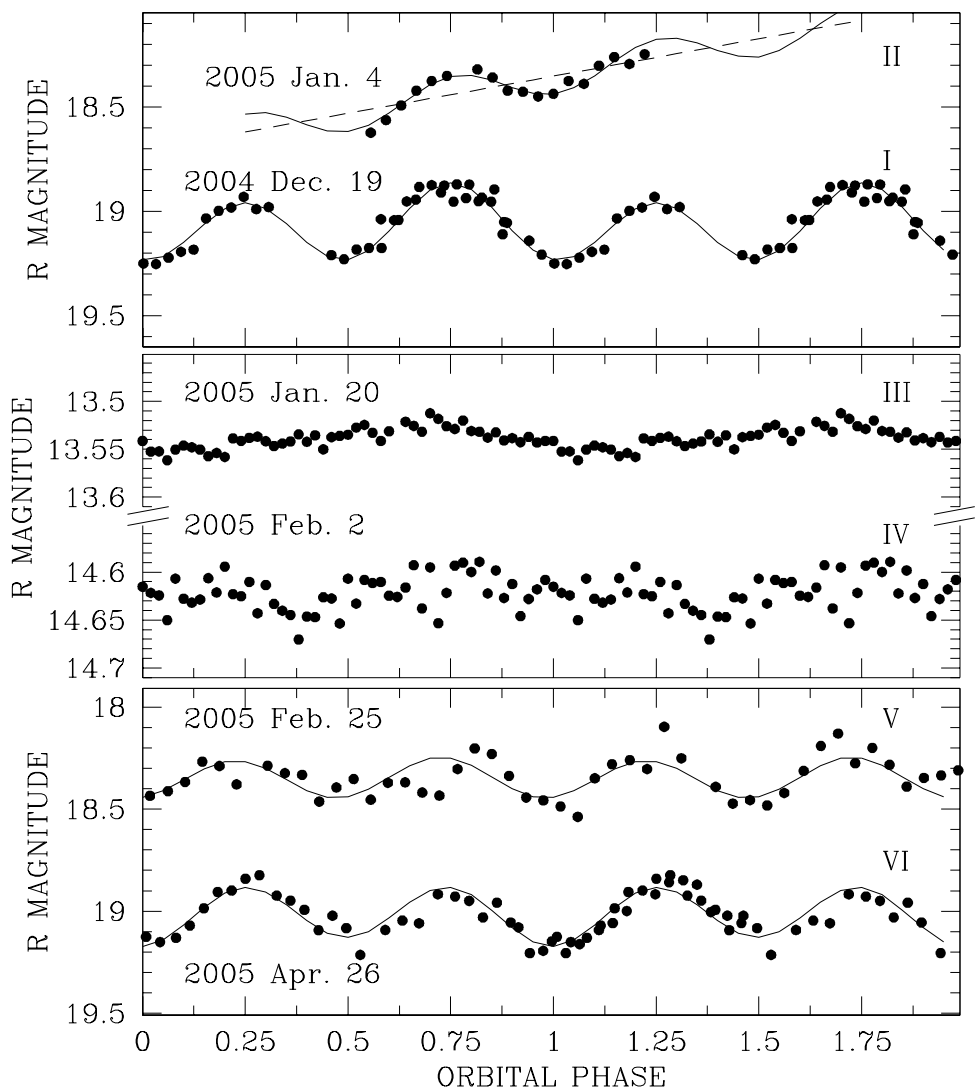}
\caption{\label{evolucion_curva} $R$--band light curves of J1118 taken during
  2004  December--  2005  April,  covering  the  different  outburst  epochs:
  Pre-outburst  quiescence  (I), rise  (II),  peak  (III),  decay (IV),  near
  quiescence (V) and post-outburst quiescence (VI).  They are phase folded on
  the ephemeris  of \citet{torres04} and  are shown twice for  clarity.  Note
  that the y-scale has been expanded when showing the January 20 and February
  2 light  curves.  December 19,  January 4, February  25 and April  26 light
  curves  have  been  fitted   with  sinusoidal  functions  to  simulate  the
  ellipsoidal modulation.}
\end{figure}

Example light curves  of J1118 at different outburst  epochs are presented in
Fig.~\ref{evolucion_curva}.  The  data have been phase folded  on the orbital
ephemeris of  \citet{torres04}.  Note the  apparent changes in  amplitude and
morphology of  the light curves as  the outburst decays. We  divide the light
curve  in 6  different stages  (see also  Fig.~\ref{outburstA}): Pre-outburst
quiescence (I),  rise (II), peak  (III), decay (IV), near-quiescence  (V) and
post-outburst quiescence (VI).\\

\noindent
(I) On 2004 December 19 (UT) J1118 was still in quiescence and displayed
the characteristic double-humped ellipsoidal modulation, with $\sim$0.16\,mag
semi-amplitude, driven by the tidally distorted secondary star.\\

\noindent
(II) On  2005 January 4, the  outburst had already started  and the $R$--band
light curve shows the combination  of the ellipsoidal modulation and a linear
rise of $0.36$\,mag/d.\\

\begin{figure}
\includegraphics[width=\columnwidth]{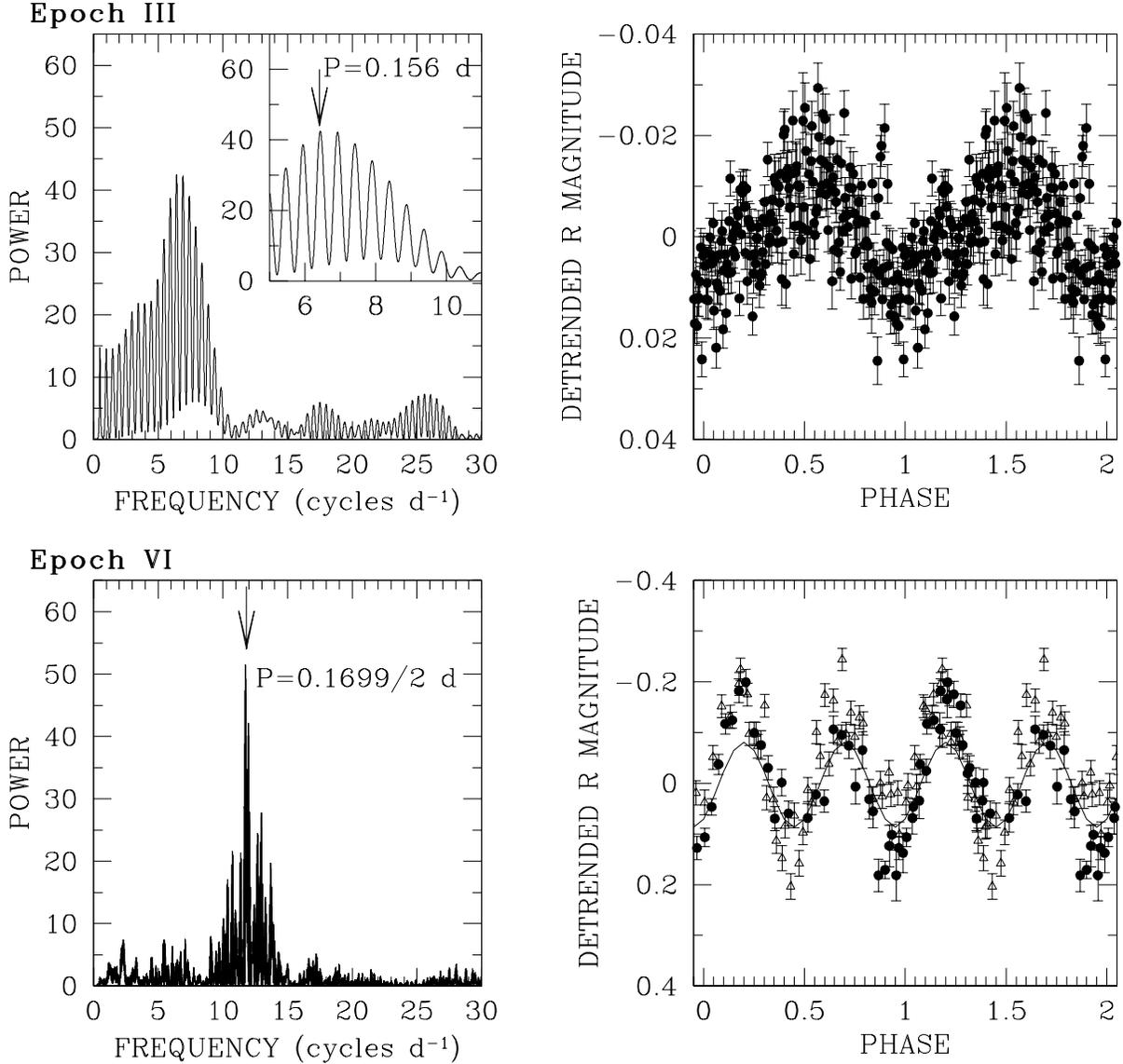}
\caption{\label{fig_optmodul}  {\em Top panels:}  The Scargle  periodogram of
 the $R$-band light  curves obtained on January 20 and  22 (near the outburst
 peak --- epoch  III) after correcting for the overall  decline and the light
 curves folded  on 0.156\,d corresponding to  the peak marked  on the Scargle
 periodogram. The data  have been averaged into 200  phase bins.  {\em Bottom
 panels:} The Scargle periodogram of  the $R$-band light curves obtained from
 March 17 to  April 26 (post-outburst quiescence --- epoch  VI) and the light
 curves obtained  on March  18 (open triangles)  and Apr 26  (filled circles)
 folded on  0.1699\,d corresponding to twice  the peak marked  on the Scargle
 periodogram. The solid line is the  sinusoidal fit of all the epoch VI data,
 simulating  the  ellipsoidal modulation.   Note  the  distortion  of the  two
 individual light  curves with respect  to the mean.  A whole cycle  is shown
 twice for continuity.}
\end{figure}

\noindent
(III) The outburst  peaks on January 14 and then the  brightness in all bands
 started  to decay  at a  moderately slow  rate of  $0.05$\,mag/d.   Near the
 outburst  peak, on  January 20  and 22,  the $R$  band light  curve  shows a
 low-amplitude modulation  superimposed on short  time scale ($\lesssim5$\,s)
 variability  likely due  to flickering.   This modulation  was  not observed
 after January 23  (alas, the $R$ band data obtained in  January 21 was badly
 affected by weather  conditions).  The same modulation was  also reported by
 \citet{chou05}  in  their $V$  band  light  curves  obtained during  January
 18--20.   This  indicates  that  either  the  low-amplitude  modulation  was
 short-lived and/or its amplitude is  diluted by the flickering or undetected
 due  to the  photometric accuracy.   We  searched for  periodicities in  the
 detrended  light  curves  of  January  20  and 22  by  computing  a  Scargle
 periodogram.  The result is shown in Fig.~\ref{fig_optmodul}.  The observing
 window produces an  alias pattern at $\sim6.5$ cycle/day  with the strongest
 peak  at $0.156  \pm 0.002$  day.  The  1-day alias  centered at  $0.169 \pm
 0.003$  day   lies  close  to   both  the  orbital  and   superhump  periods
 \citep[see][]{uemura00,zurita02}.   Although   the  observed  modulation  is
 probably related  to the orbital motion,  the poor data  sampling impedes an
 accurate  period determination.  In  Fig.~\ref{evolucion_curva} we  show the
 light  curve of  January 20  light curve  folded on  the orbital  period and
 averaged into 50  phase bins, although the cleanest  modulation is seen when
 folding   the   data    on   $0.156$   day   (see   left    top   panel   of
 Fig.~\ref{fig_optmodul}).\\

\noindent
(IV) From the beginning of February the light curve began a moderately abrupt
fall.  We  estimated a  rate  of $\sim0.15$\,mag/d  in  the  $R$ band and  a
smoother slope  of $0.10$\,mag/d  in the infrared  bands.  On February  2, no
modulation     is    detected     but     only    considerable     flickering
($\sigma_{\mathrm{m}}\sim$0.02\,mag), higher  than found at the  peak of the
outburst.\\

\noindent
(V) At  the end of  February, about 60  days after the outburst  onset, J1118
settled    into    a     near-quiescence    level    at    $R$=18.35$\pm$0.02
(Fig.~\ref{evolucion_curva}).  On  February 25 the light  curve is consistent
with a (distorted) double-humped ellipsoidal modulation with a semi-amplitude
of  $\sim$0.10 mag.   We  also note  that  the amplitude  of the  ellipsoidal
modulation is  lower at  near-quiescence than at  true quiescence  before and
after outburst.   This is  what we  would expect if  the contribution  of the
accretion  disk to  the total  light is  higher in  February 25  (V)  than in
December 19  (I) and April  26 (VI).  Assuming  that the decrease in  flux is
solely due to  the accretion disk light fading and  that the disk contributed
$\sim$55\%  to  the  total  quiescent $R$--band  light  \citep{torres04},  we
estimate a relative  contribution of the accretion disk  of $\sim$77\% during
epoch V. \\

\noindent
(VI) Two months later, J1118  faded another 0.6\,mag and reached $R\sim$19 on
March 17.   The system remained  at this level  from there on,  suggesting it
reached true quiescence.  At the  same epoch we measured the following colors
in the infrared: $K_\mathrm{s}$=16.66$\pm$0.07, $H-K_\mathrm{s}$= 0.8$\pm$0.2
and $J-K_\mathrm{s}$=1.1$\pm$0.1.  The  $J$ and $K_\mathrm{s}$ magnitudes are
consistent   with  those  observed   by  \citet{mikolajewska05}   during  the
pre-outburst  quiescence (our  epoch  I).  The  colours $H-K_\mathrm{s}$  and
$J-K_\mathrm{s}$ are  much redder  than expected  for a later  K or  early MV
secondary star suggesting an additional contribution. \\

The Scargle periodogram of the  epoch VI $R$--band lightcurves (March 17, 19,
20, 23  and April 22, 26)  shows a strong peak  centered on $0.0845\pm0.0005$
day (see left bottom  panel of Fig.~\ref{fig_optmodul}), consistent with half
the orbital period $P_\mathrm{orb}=0.1699$ day.  Also, these light curves are
distorted likely due to the presence  of a superhump wave, as already noticed
in   the  near-quiescence   state   at   the  end   of   the  2000   outburst
\citep{zurita02}. In  contrast, on  December 19 and  January 4 (epochs  I and
II),  there  is  no  evidence  of superhumps.   This  suggests  that,  during
outburst, the disk  expands to the 3:1 (or 2:1) resonance  radius and is then
forced  to precess  by  tidal  perturbations caused  by  the secondary  star.
Afterwards,  it starts to  shrink although  at early  quiescence it  is still
large enough to produce superhump  waves.  Finally, some time later, the disk
radius becomes shorter than the resonance radius and superhumps disappear.

\section{An optical precursor to the X-ray outburst?}
\label{delay}

There has  been some evidence  of optical/infrared outbursts  starting before
the  X-ray outbursts  in  some XRTs:  GS  1124-684 \citep{dellavalle91},  GRO
J1655--40  \citep{orosz97},  GRO  J0422+32 \citep{castrotirado97},  V404  Cyg
\citep[and  references  therein]{chen97},  Aql X-1  \citep{shahbaz98},  J1118
during  the  main  2000   (March)  outburst  \citep{wren01}  and  4U\,1543-47
\citep{buxton04}.   Also, in dwarf  novae the  UV rise  has been  observed to
start several  hours after the optical  outburst \citep[e.g.,][and references
therein]{warner95}.   In the framework  of the  {\em Disk  Instability Model}
\citep[see  e.g.,][]{cannizzo95},  the  X-ray   (or  UV)  delay  suggests  an
`outside-in'  disturbance of  the accretion  disk.  Once  the  instability is
triggered in  the outer regions,  a heating front propagates  inwards turning
the disk from  the cold (quiescent) state to a hot  state. Hence the outburst
is first noticed in  the optical and then in X-rays (or  UV).  The time scale
of the lags can be explained assuming the accretion disk is truncated at some
inner  radius.  The heating  front stops  when it  arrives at  the truncation
radius, but  the inner edge of the  disk moves towards the  compact object on
the viscous  time scale,  longer than the  front propagation time.   The ADAF
model  can offer  a  natural explanation  for  the disk  truncation in  XRTs.
Hence, it  has been  proposed that  the disk inner  truncation radius  can be
estimated by measuring the X-ray to optical delay \citep{hameury97,wren01}.\\

To take advantage  of the fact that our optical  observations covered part of
the rise  phase (see Fig.~\ref{outburstA}), we inspected  whether the optical
and  X-ray  outbursts  were simultaneous  or  if  one  lags the  other.   The
time-resolved   optical  light   curve   taken  on   2005   January  4   (see
Fig.~\ref{evolucion_curva}) shows  a linear rise.  We  therefore estimate the
starting   time  for   the  optical   rise  to   be   $t_\mathrm{opt}  \simeq
2453375.1\pm0.1$ (HJD) from  a linear fit to this  curve.  A more problematic
issue is  to determine the  starting time of  the X-ray outburst.   The X-ray
light curve  from ASM does  not provide any  useful information below  the 10
mCrab  sensitivity level and  hence extrapolation  is required.   We estimate
that  the X-ray  outburst starting  time is  consistent with  $t_\mathrm{X} =
2453379.4  \pm  0.7$  (HJD),  where  the associated  uncertainty  quotes  the
differences between the several  extrapolations we performed.  This implies a
$\sim4$-day  lag  between the  onset  of  the  X-ray and  optical  outbursts.
Unfortunately, the  ASM sensitivity  level is well  above the  quiescent flux
level  making  it  very  likely  that  any  extrapolation  overestimates  the
delay. Besides,  most of  the sources  detected by the  ASM need  to brighten
significantly  above  quiescence to  be  detected  \citep[see e.g.][for  more
details  on this  concern]{homan05}. Therefore  the true  start of  the X-ray
outburst  could be  earlier  than  $t_\mathrm{X}$.  This  fact  forces us  to
conclude that the  4-day lag is just an  upper limit as it is  the 10-day lag
estimated by  \citet{wren01}.  In short,  although we can draw  a qualitative
picture of the  evolution of the accretion disk in  J1118 after the outburst,
it is not possible to estimate a reliable pre-outburst truncation radius from
the X-ray delay using ASM data alone.

\section{The spectral energy distribution}

\begin{figure}
\includegraphics[angle=0,scale=0.8]{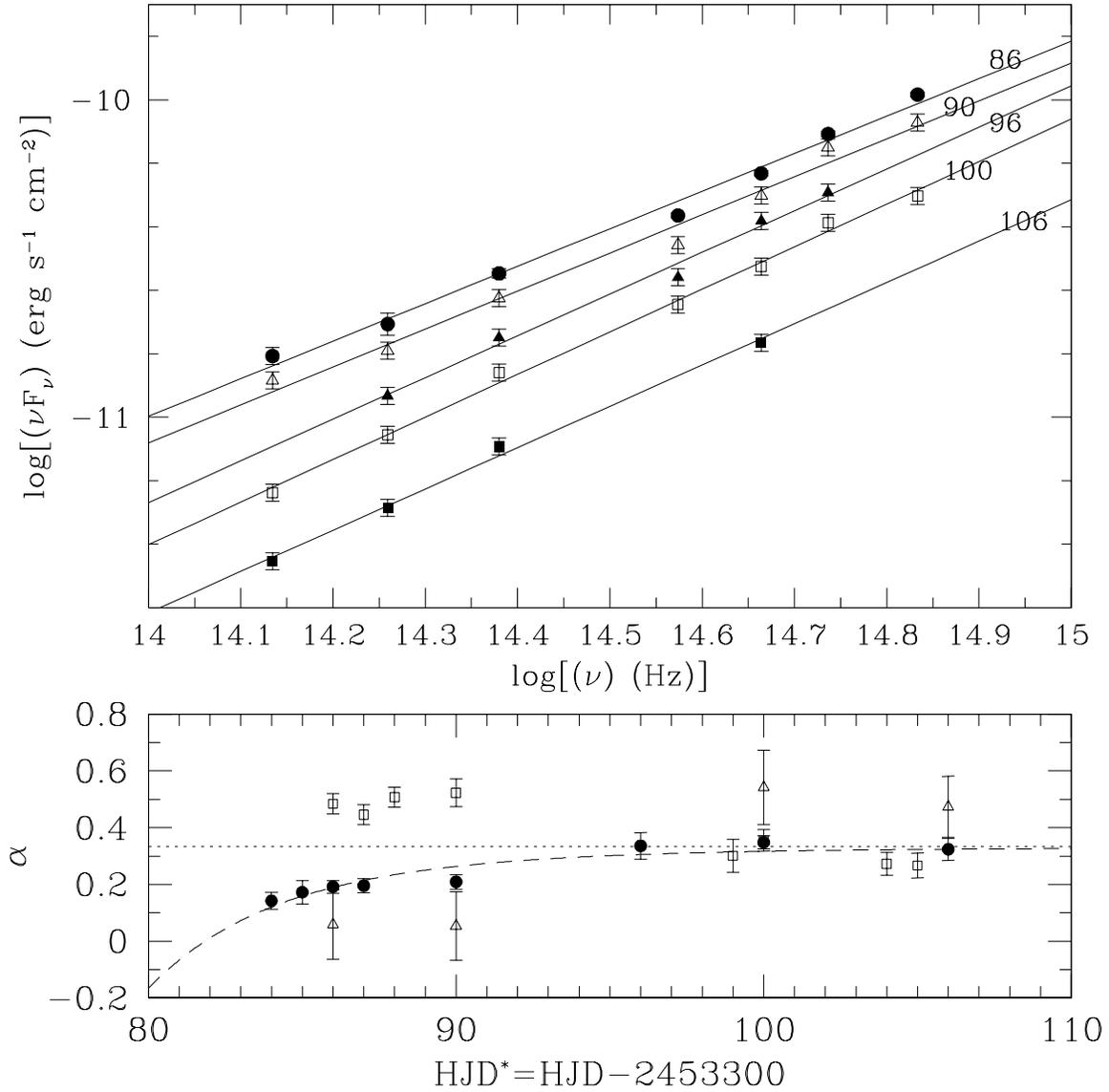}
\caption{{\em Top  panel:} Evolution of  the optical-infrared SEDs  for J1118
  through the outburst.  Different  symbols are used to distinguish alternate
  epochs   and   solid   lines   are   power   law   fits.    Numbers   label
  HJD$^{*}$=HJD-2453300   (note  that   the  optical   outburst   started  at
  HJD$^{*}$=75).    {\em   Bottom   panel:}   Evolution  of   the   power-law
  index---$\alpha$  (where $F_{\nu}\propto\nu^{\alpha}$).   Open  squares are
  the indices obtained from the  optical SEDs (from log$\nu$=14.54 to 14.83),
  open triangles correspond to the  infrared SED fits (from log$\nu$=14.13 to
  log$\nu$=14.38) whereas  filled circles are  the indices obtained  from the
  optical-infrared SED fits (from  log$\nu$=14.13 to 14.83).  The dotted line
  corresponds to the canonical $\nu^{1/3}$ value for a steady state viscously
  heated disk. The  dashed line shows the exponential  fitting to the indices
  obtained from the whole wave-length range SEDs.}
\label{seds}
\end{figure}

We  constructed  the  spectral  energy  distributions  (SEDs)  for  different
outburst epochs  through the outburst.   The magnitudes were  first corrected
for  interstellar  extinction using  $E(B-V)$=0.013  \citep{hynes00} and  the
reddenings  tabled  in \citet{rieke85},  although  this  makes  only a  small
difference in  the dereddened  magnitudes because the  extinction is  so low.
Our outburst SEDs  are shown in Fig.~\ref{seds}.  Here  we have excluded some
nights for clarity to avoid  duplication.  All SEDs are quasi-power-laws with
$F_{\nu}$ increasing  toward the blue.   We have performed power-law  fits to
the  optical SEDs  only (from  log\,$\nu$=14.54  to 14.83),  to the  infrared
magnitudes  alone (from  log\,$\nu$=14.13 to  14.38)  and then  to the  whole
wavelength range  (from log\,$\nu$=14.13 to 14.83).  The  spectral indices we
found  have   been  plotted  as  a   function  of  time  and   are  shown  in
Fig.~\ref{seds} (bottom panel), where $F_{\nu}\propto\nu^{\alpha}$.\\

When considering the optical data alone (open squares in Fig.~\ref{seds}) the
source appears  steeper (bluer) than  the canonical $\alpha=1/3$ of  a steady
state viscously  heated disk, being  $\alpha=0.49\pm0.04$ at the peak  of the
outburst  (from HJD$-$2453300=85  to 90).  During  the decay  phase (IV),  it
becomes  optically softer  with $\alpha\sim0.25$.   This trend  is consistent
with the cooling of the  optically bright regions.  However, when fitting the
infrared alone the  trend is reversed showing that  the infrared evolves very
differently (open triangles) than the optical. The joint fit (filled circles)
is  the interplay  between these  components and  shows an  exponential trend
becoming  bluer throughout the  outburst.  At  the decay  phase the  index is
consistent  with  $\alpha=1/3$.  We  will  discuss  the  implications of  the
different SEDs in the next section.

\section{Summary and discussion}

We present  the follow-up of  the new outburst  of the halo black  hole X-ray
transient  XTE  J1118+480.   We   estimate  a  total  outburst  amplitude  of
$5.7\pm0.1$  in  the   $R$  band  and  $\sim5$\,mag  in   the  $J$,  $H$  and
$K_\mathrm{s}$-bands.  The shape of the  light curve is remarkably similar to
the mini-outbursts observed in  GRO J0422+32 and XTE J1859+226.  Furthermore,
$L_\mathrm{X}$(1-10   keV)$/L_\mathrm{opt}\sim$5  similar   to  mini-outburst
episodes, whereas typically this ratio is $\sim$500 in normal X-ray transient
outbursts.\\

Quiescent XRTs  are often explained  by assuming an accretion  disk extending
down to a  certain transition radius while the inner volume  is filled with a
hot  ADAF  \citep[e.g.,][]{narayan96}.    This  model  predicts  that,  after
outburst,  the  inner  disk  edge  will  move  farther  inward.  Because  the
efficiency  of energy  release in  the  ADAF region  is very  low, the  X-ray
outburst starts when  the densest parts of the disk  can penetrate far enough
in  to  allow  an  efficient  transformation  of  gravitational  energy  into
radiation.  The outer disk is also  expected to change.  In the course of the
outburst,  matter diffuses inwards  closing the  disk while  angular momentum
transferred to  the outer parts forces  the outer radius to  increase.  If it
reaches the 3:1 resonance radius,  the disk will precess by tidal interaction
with  the  secondary   star  \citep[e.g.,][]{whitehurst91}  and  a  superhump
modulation will be presumably visible.  This  was the case in March 2000 when
the change in the outer disk radius  could be measured from the change in the
superhump  period  at  different  stages of  the  outburst  \citep{zurita02}.
Equally, in  2005 we  found a 0.02-mag  amplitude variation during  the early
decline and also a  distorted double-humped ellipsoidal modulation during the
near-quiescence  level and  true  quiescence.  This  suggests  that the  disk
expanded after  outburst to the 3:1  resonance disk radius  where it remained
during the early phases of quiescence.\\

Recently,  the  LH state  of  X-ray binaries  has  been  associated with  jet
 activity.   In   some  cases  a   jet  like  structure  has   been  resolved
 \citep[e.g.,][]{mirabel92}.  When jets cannot  be directly imaged, a flat or
 even inverted radio  spectrum is often considered to  be a typical signature
 of jet emission \citep{fender01}.  However, it is also apparent that the jet
 contributes  outside the  radio  band.  The  synchrotron  spectrum, that  is
 thought to be the jet signature, is  frequently seen at radio but also up to
 higher frequencies in the infrared and possibly in the optical.  In the case
 of J1118,  the SED from  radio to X-rays  during the 2000 outburst  has been
 explained  as  a combination  of  synchrotron radiation  from  a  jet and  a
 truncated  optically  thick  disk \citep{hynes00,markoff01,yuan05b}  whereas
 models assuming  ADAF alone \citep{mcclintock01b,esin01}  underestimated the
 optical and the  infrared fluxes.  The spectrum from infrared  to UV is flat
 \citep[$F_{\nu}\sim$constant;  ][]{hynes00}  although  the optical  spectrum
 alone has blue continuum slopes of $\alpha=1/3$ as expected for an optically
 thick accretion disk \citep{dubus01,torres02}.\\

The  SEDs  during the  2005  outburst  exhibit  quasi-power-law spectra  with
$\alpha$ softening from  $\sim$0.49 during the peak of  the X-ray outburst to
$\sim$0.25 during the decay phase.   However, when fitting the infrared alone
we find a  flat spectrum with $\alpha=0.1\pm0.1$ at  the outburst peak.  This
difference between the  optical and the infrared SEDs,  more important at the
outburst  peak,  suggests that  the  infrared  is  dominated by  a  different
component (a jet?)  whereas in the  optical we are presumably seeing the disk
evolution. The very flat infrared SED ($F_{\nu}\sim$constant) could naturally
be  interpreted  as  a  mixture  of  an optically  thick  disk  spectrum  and
flat-spectrum emission,  possibly synchrotron.   Linear fits to  optical SEDs
have also been  performed for other short-period black  hole XRTs in outburst
\citep[see  the   compilation  by][]{hynes05}.  The  optical   SEDs  for  the
\citet{hynes05} relatively uniform  set, exhibit quasi-power-law spectra with
$\alpha$  ranging  between  0.5  and  1.5, all  steeper  than  the  canonical
$F_{\nu}\propto\nu^{1/3}$.   Two  of the  sources  among the  \citet{hynes05}
sample   (GRO    J0422+32   and    XTE   J1859+226)   were    identified   by
\citet{brocksopp04} as hard sources (XTE  J1859+226 is hard at least early in
the  outburst).  The  spectra  of both  sources  exhibit a  quasi-exponential
softening throughout the outburst whereas  the other systems exhibit no clear
trends. However,  very little data are  available for these two systems. 

Our  data clearly  demonstrate the  added value  of extending  the wavelength
range into  the near-infrared.  We  were able to witness  additional spectral
components  that   show  a   different  trend  during   the  course   of  the
outburst. Extending  the wavelength coverage even further  would have allowed
for  a  more  quantitative  comparison  with  proposed  descriptions  of  the
accretion   flows  near   compact  objects.    Looking   ahead,  simultaneous
multi-wavelength observations from X-rays through  to radio will enable us to
validate the  interplay between disks  and jets.  J1118 remains  an excellent
target for multi-wavelength studies, which  needs to be exploited with future
and present facilities as we look forward to its next outburst.

\section{Acknowledgments}

MAPT thanks the observers at the  1.5m telescope at FLWO (in particular Perry
Berlind and Mike Calkins) for helping during the remote observations with the
1.2m. This work was supported by NASA LTSA grant NAG-5-10889. DS acknowledges
a  Smithsonian   Astrophysical  Observatory  Clay   Fellowship.   The  Peters
Automated  Infrared   Imaging  Telescope   (PAIRITEL)  is  operated   by  the
Smithsonian Astrophysical Observatory (SAO) and  was made possible by a grant
from the Harvard University Milton  Fund, the camera loan from the University
of Virginia, and the continued support of the SAO and UC Berkeley.

\end{document}